\documentclass[aps,showpacs,preprint,superscriptaddress]{revtex4}
\usepackage{amsmath,amssymb,amsfonts}
\usepackage[english]{babel}
\usepackage{graphicx}
\usepackage{dcolumn}
\usepackage{bm}
\usepackage{color}

\begin{document}

\title{Dynamical constants of structured photons with parabolic-cylindrical symmetry}

\author{B. M. Rodr\'{\i}guez-Lara}
\affiliation{Instituto de F\'{\i}sica, Universidad Nacional Aut\'{o}noma de M\'{e}xico,
Apdo. Postal 20-364, M\'{e}xico D.F. 01000, M\'{e}xico.}

\author{R. J\'auregui}
\affiliation{Instituto de F\'{\i}sica, Universidad Nacional Aut\'{o}noma de M\'{e}xico,
Apdo. Postal 20-364, M\'{e}xico D.F. 01000, M\'{e}xico.}

\date{\today}

\begin{abstract}
Electromagnetic modes with parabolic-cylindrical symmetry and their dynamical
variables are studied both in the classical and quantum realm. As a result, a new
dynamical constant for the electromagnetic field is identified and
linked to the symmetry operator which supports it.
\end{abstract}


\maketitle

\section{Introduction}

The dynamical variables of the electromagnetic (EM) field  define
its mechanical identity and are essential for understanding the
effects of the field on charged particles. As a consequence,
finding the natural dynamical variables of the EM field and their
relations to mode properties has been directly linked to the
development of classical and quantum EM theory. Historically, the
photon concept emerged from suggesting a definite relationship
between the energy (linear momentum) of a photon and the frequency
$\omega$ (wave vector  $\vec k$) of plane waves. Similarly, the
relationship between the angular momentum of EM waves and their
polarization is very important for understanding, e. g., atomic
processes mediated by photons.

Photons associated to EM modes with non Cartesian symmetries are characterized by
 sets of  dynamical constants different from those of plane waves. An example
corresponds to circular-cylindrical EM waves known as Bessel modes \cite{Durnin}
or their paraxial analogue, i. e., Laguerre-Gaussian beams \cite{Allen}.
Bessel photons carry a well defined orbital angular momentum \cite{Allen, deBroglie}
proportional to the winding number $m$ of their vortices \cite{Bouchal, Jauregui2005}.
Another example corresponds to Mathieu modes which exhibit elliptical-cylindrical
symmetry. Mathieu photons carry constant values for the
balanced composition of the orbital angular momentum with respect to the foci
of the elliptical coordinate system \cite{Rodriguez2008}.

The purpose of this paper is to analyze the mechanical properties of
 the EM modes for the last coordinate system with translational
 symmetry along an axis known to have separable analytical solutions, i. e., the
 parabolic-cylindrical coordinate system. The solutions of the corresponding
 wave equation can be expressed in terms of Weber functions, giving this name to the
 EM modes. We show that a balanced composition of a component of the lineal
  momentum with a component of the angular momentum
is a  natural dynamical variable for these modes. Weber photons
carry a well defined value of this variable.

 Weber beams of zero-order have already been experimentally generated by means
 of a thin annular slit modulated by the proper angular spectra \cite{Lopez2005}.
 This setup was conceived as a variation of that originally used by Durnin et al.\cite{Durnin}
 for generating Bessel beams. Higher order Weber beams can also be produced by
 holograms encoded on plates \cite{Lopez2005} or in spatial light modulators \cite{Davis}.

{\bf Parabolic-cylindrical coordinates.}

The parabolic-cylindrical coordinate system $(u,v,z)$ is defined by the
transformations \cite{Lebedev1966}
\begin{equation}
x + i~y = \frac{1}{2} \left( u + i~v \right)^2,\quad z=z
\end{equation}
where $x$, $y$ and $z$ are the well known Cartesian coordinates, and
$u\in \left(-\infty, \infty \right)$ and $v \in \left[ 0, \infty \right)$.
 Surfaces of constant $u$ form half confocal parabolic cylinders
that open towards the negative $x$ axis, while the surfaces of
constant $v$ form confocal parabolic cylinders that open in the
opposite direction. The foci of all these parabolic cylinders
are located at $x$=0 and $y$=0 for each $z$ value. The scaling factors associated to
$u$ and $v$ are $h_{u} = h_{v} = h = \sqrt{u^2 + v^2}$.
In the following, the notation $\hat{e}_{x}$ represents the unitary vector
related to a given coordinate $x$, the shorthand notation $\partial_{x}$ is
used for partial derivatives with respect to the variable $x$, and
$\partial_0=:\partial_{ct} = \frac{1}{c} \partial_{t}$ with $c$ the
velocity of light in vacuum.

\section{A parabolic scalar field and its dynamical variables}

The scalar wave equation has separable solutions
invariant under axial propagation,
\begin{equation}
\nabla^2 \Psi = \partial_{ct}^2 \Psi, \quad \Psi (\vec{r},t) =
 \psi(\vec{r}_{\perp}) e^{i \left( k_{z} z - \omega t \right)}/\sqrt{2\pi}\label{eq:total}
\end{equation}
in four coordinate systems: Cartesian, circular-, elliptic- and
parabolic-cylindrical coordinates.
 For parabolic-cylindrical symmetry  Helmholtz equation reads
\begin{equation}
\left[ h^{-2} \left( \partial_{u}^{2} + \partial_{v}^{2} \right) + k_{\perp}^2 \right] \psi(u,v) = 0,  \quad k_{\perp}^2 = k^2 - k_{z}^2, \label{eq:Helmholtz}
\end{equation}
where the real constants $k= \omega / c$ is the magnitude of the
wave vector for a given frequency $\omega$, $k_z$ its axial
component and $k_\perp$ its perpendicular component. If $\psi(u,v)
= U(u) V(v)$
\begin{eqnarray}
\left(\partial_{u}^{2} + k_{\perp}^2 u^2 - 2 k_{\perp} a \right) U(u) = 0, \label{eq:weberu} \\
\left(\partial_{v}^{2} + k_{\perp}^2 v^2 + 2 k_{\perp} a \right) V(v) = 0\label{eq:weberv},
\end{eqnarray}
with $2k_\perp a$ the separation constant. These equations are
known as parabolic cylinder or Weber differential equations.
Solutions for this differential set can be expressed as Frobenius
series, parabolic cylinder functions, Whittaker functions, Hermite
functions, and others \cite{Abramowitz1972,
Lebedev1965,miller1974}. Here, the solutions are expressed in
terms of confluent hypergeometric functions of the first
kind,$_{1}F_{1}$:
\begin{equation}
U_{p,k_\perp,a}(u) = s_p \zeta_u^{\frac{n_p-1}{4}} e^{-i\frac{\zeta_u}{2}} ~_{1}F_{1} (\frac{n_p}{4} - i
 \frac{a}{2}, \frac{n_p}{2}; i \zeta_u),
\end{equation}
\begin{equation}
V_{p,k_\perp,a}(v) = s_p\zeta_v^{\frac{n_p-1}{4}} e^{-i\frac{\zeta_v} {2}} ~_{1}F_{1} (\frac{n_p}{4}+ i\frac{a}{2},\frac{n_p}{2}; i \zeta_v ),
\end{equation}
with $\zeta_u = k_\perp u^2$, $\zeta_v = k_\perp v^2$,
$n_e=1$  ($n_o=3$) for even (odd) parity functions:
$U_e(-u) = U_e(u)$  ($U_o(-u) = -U_o(u)$). The normalization factors are taken as
$$s_e = \frac{\sqrt{ \pi\sec(ia\pi)}}{\vert \Gamma(3/4-ia/2)\vert},\quad
  s_o = \frac{\sqrt{2\pi\sec(ia\pi)}}{\vert \Gamma(1/4-ia/2)\vert}.$$
These expressions reduce directly to the Frobenius series,
presented in Ref.~\cite{Abramowitz1972} and used in
Ref.~ \cite{Bandres2004} to introduce parabolic optical wave
fields in the paraxial regime. In order to guarantee that these scalar fields
vanish when the absolute values of the coordinate variables tend to infinity,
 $a$ must be real \cite{Lebedev1965}.
 The set $\{ \Psi_{p,\kappa},\kappa=(k_{z},\omega,a)\}$ is complete and orthogonal.
 Each function $\Psi_{p,\kappa}$ satisfies the eigenvalue equations,
\begin{eqnarray}
{\mathfrak{P}}_u~\Psi_{p,\kappa}(u,v,z,t) &\doteq& \Psi_{p,\kappa}(-u,v,z,t)\nonumber\\
                          &=& (-1)^p\Psi_{p,\kappa}(u,v,z,t) \label{eq:par}\\
-i\partial_z~\Psi_{p,\kappa}(u,v,z,t) &=&  k_z\Psi_{p,\kappa}(u,v,z,t)\label{eq:pz}\\
 i\partial_{t}~\Psi_{p,\kappa}(u,v,z,t) &=& \omega \Psi_{p,\kappa}(u,v,z,t)\label{eq:time}\\
2\mathbb{A}~\Psi_{p,\kappa}(u,v,z,t) & \doteq & \left\{\frac{ v^2}{h^2} \partial_{u}^{2} - \frac{u^{2}}{h^2} \partial_{v}^2 \right\}\Psi_{p,\kappa}(u,v,z,t)\nonumber\\
 &=&2k_{\perp} a~\Psi_{p,\kappa}(u,v,z,t)\label{eq:A}
 \end{eqnarray}
The operator $\mathbb{A}$ is directly identified as a generator of
the balanced composition of a rotation around the z-axis and
translations along the $y$-axis since
\begin{equation}
\mathbb{A} =(1/2) \partial_{x} +  y \partial_{xy}^2 -  x \partial_{y}^2 =
 \left( l_{z} p_{y} + p_{y} l_{z} \right)/2
\end{equation}
where $l_z =-i (\vec r\times\vec\nabla)_z$, and $p_y =-i
\partial_y$. For scalar fields and spacetime continuous
symmetries, the generators of infinitesimal transformations turn
out to be good realizations of the corresponding dynamical
operator. In that sense, $\mathbb{ A}$ can be related to the
product of the z-component of the angular momentum and the
y-component of the linear momentum. Then, the eigenvalue equation
Eq.(\ref{eq:A}) means that the scalar field
$\Psi_{p,k_z,\omega,a}$ carries a well defined value of that
momenta product.

Travelling scalar Weber modes are defined by
\begin{equation}\label{eq:trav}
\tilde\Psi_{\pm}\kappa(\vec r,t)
 =(\psi_{e,\kappa}(u,v) \pm i \psi_{o,\kappa}(u,v))(e^{i(k_zz-\omega t)})/\sqrt{2\pi}, \end{equation}
and these modes  are orthonormal \cite{miller1974}
\begin{equation}
\int_{\mathbb{R}^3}\tilde\Psi_{\pm,\kappa^\prime}^{\ast}(\vec r)\tilde\Psi_{\pm,\kappa^\prime}(\vec r)
 = \frac{2\pi}{k_\perp} \delta(k_z-k_z^{\prime})
\delta(k_\perp - k_\perp^\prime) \delta(a-a^{\prime}).\nonumber
\end{equation}
\begin{figure}
\includegraphics[width= 0.5 \textwidth]{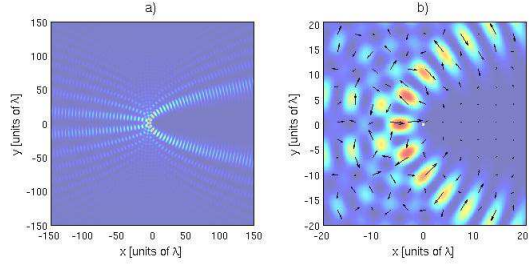}
\caption{\label{fig:Fig1}(Color Online) Sample of (a) transverse
intensity and (b) polarization  structure of an odd EM TE Weber
field.  They correspond to the eigenvalue $a=-2$, $k_z =0.995
\omega/c$ and the unit of length is taken as the wave length.}
\end{figure}

\section{Parabolic-cylindrical EM modes}

In Coulomb gauge, any given solution for the vector electromagnetic field,
$\vec{A}$, can be written as a superposition of modes related to a complete
set of solutions for the scalar wave equation, $\Psi$, which are
identified as Hertz potentials \cite{Stratton1941}.
For EM fields in parabolic-cylindrical coordinates having a well defined
behavior under ${\mathfrak{P}}_u$
\begin{equation}
\vec A_{p,\kappa} = \mathcal
{A}^{(TE)}_{p,\kappa} \vec{\mathbb{M}} \Psi_{p,\kappa} + \mathcal {A}^{(TM)}_{p,\kappa} \vec{\mathbb{N}}
\Psi_{p,\kappa} \label{eq:parabolic}
\end{equation}
with the vector operators given by the expressions
\begin{equation}
\vec{\mathbb{M}} = \frac{ \partial_{ct}}{h} \left( \hat{e}_{u} \partial_{v} -
\hat{e}_{v} \partial_{u}\right),  \vec{\mathbb{N}} = \frac{
\partial_{z}}{h} \left( \hat{e}_{u} \partial_{u} + \hat{e}_{v} \partial_{v}\right) -
\hat{e}_{z} \nabla_{\perp}^{2}.
\end{equation}
The constants $\mathcal{A}^{(TE)}_{p,\kappa}$ and
$\mathcal{A}^{(TM)}_{p,\kappa}$ are proportional to the amplitude
of the transverse electric (TE) and transverse magnetic (TM) EM
fields, as can be directly seen from their connection with the
associated electric and magnetic fields, $\vec{E} = -
\partial_{ct}\vec{A}$ and $\vec{B} = \vec{\nabla} \times \vec{A}$
yielding
\begin{eqnarray}
\vec{E}_{p,\kappa} &=& - \mathcal{A}^{(TE)}_{p,\kappa}\partial_{ct}\vec{\mathbb{M}}
 \Psi_{p,\kappa} - \mathcal{A}^{(TM)}_{p,\kappa}\partial_{ct}\vec{\mathbb{N}} \Psi_{p,\kappa}, \nonumber\\
\vec{B}_{p,\kappa} &=&\quad \mathcal{A}^{(TE)}_{p,\kappa}\partial_{ct}\vec{\mathbb{N}}
\Psi_{p,\kappa} - \mathcal{A}^{(TM)}_{p,\kappa}\partial_{ct}\vec{\mathbb{M}} \Psi_{p,\kappa}.
\end{eqnarray}
Similar expressions can be written for the travelling  EM modes
associated to Eq.~(\ref{eq:trav}).
 The intensity and polarization structure of a EM Weber beam are illustrated in
 Fig.~\ref{fig:Fig1}. The instantaneous
electric field orientation has a nontrivial structure and, as a function of
time, ${\bf E}$ preserves its direction while its magnitude oscillates at each point.

\section{Dynamical constants for parabolic-cylindrical EM modes}
Given a symmetry generator Noether theorem is usually applied to the field Lagrangian density
\begin{equation}
{\mathfrak L} = \frac{1}{8\pi}\sum_{\mu,\nu=0}^3(\partial_\mu A_\nu - \partial_\nu A_\mu)(\partial^\mu A^\nu - \partial^\nu A^\mu),
\end{equation}
in order to find a dynamical constant for the electromagnetic
field.
This theorem states that \cite{Bogoliubov1980},
if under an infinitesimal
transformation of the space coordinates $x_\mu\rightarrow x_\mu +\sum_\rho X^\rho_\mu \delta\omega_\rho$ and
the field $A_\mu \rightarrow   A_\mu + \sum_\rho\Phi^\rho_\mu \delta\omega_\rho,$
the Lagrangian is left invariant, then the current
\begin{eqnarray}
\Theta^\nu_\rho  &=& \Xi^\nu_\rho + \Lambda^\nu_\rho  \\
\Xi^\nu_\rho &=&- \sum_{\lambda}\frac{\partial{\mathfrak L}}
{\partial (\partial_\nu A_{\lambda})}\Phi_{\lambda \rho} \label{eq:intrinsec}\\
\Lambda^\nu_\rho &=& \sum_{\lambda,\sigma}
\frac{\partial{\mathfrak L}}{\partial( \partial_\nu A_{\lambda})}
X^\sigma_\rho A_{\lambda,\sigma}- {\mathfrak L} X_\rho^\nu
\label{eq:extrinsec}
\end{eqnarray}
has a null divergence. As a consequence,
$\Theta^0_\rho$ defines the density of a dynamical variable whose integrated value over a volume
can change only due to the flux of the current $\Theta_\rho^i$ through the surface that
delimits the volume.

Since we are working with EM modes that have a particular symmetry
the subset of all possible transformations whose generators are directly identified
from the eigenvalue equations~(\ref{eq:par}-\ref{eq:A}) is particularly relevant.
Under an infinitesimal translation along the main direction of propagation
$\delta z$ or a time translation $\delta t$, the EM field changes according to
the expressions $A_\mu\rightarrow A_\mu +\partial_z A_\mu \delta z$ or
$A_\mu \rightarrow A_\mu + \partial_{t} A_\mu \delta t$, respectively.
This is reflected in the fact that the field momentum-like variable
\begin{equation}
P^{(i,p,\kappa,p^\prime,\kappa^\prime)}_{z} = \frac{1}{4\pi c}\int d^3x
(\vec E^{(i)}_{p,\kappa}\times \vec
B^{(i)}_{p^\prime,\kappa^{\prime}})_{z},i=TE,TM,
\label{eq:momentum}
\end{equation}
is independent of time if the integration is taken over the whole space.
Similarly, the  energy-like integral
\begin{equation}
\mathcal{E}^{(i,p,\kappa,p^\prime,\kappa^{\prime})}= \frac{1}{4\pi}\int d^3x
\left[{\vec{E}}^{(i)}_{p,\kappa} \cdot
\vec{E}^{(i)}_{p^\prime,\kappa^{\prime}}
+\vec{B}_{p,\kappa}^{(i)} \cdot
\vec{B}_{p^\prime,\kappa^{\prime}}^{(i)}\right]\label{eq:energy}
\end{equation}
is also constant. In fact, $
P^{(i,p,\kappa,p^\prime,\kappa^\prime)}_{z}$ and $\mathcal
{E}^{(i,p,\kappa,p^\prime,\kappa^\prime)}$ are proportional to
each other with $k_z/\omega$ the constant of proportionality.
Notice that, in both cases, the dynamical constant can be inferred
from the factor $\Xi^\nu_\rho$ defined in Eq.~(\ref{eq:intrinsec})
up to a term proportional to the divergence of a vector field, for
instance
\begin{equation}
\Xi^0_{j}=\frac{1}{4\pi} \sum_i E_i \partial_j A_i=\frac{1}{4\pi}
(\vec E\times\vec B)_j - \sum_i
\partial_i(E_iA_j).
\end{equation}
$\Xi^\nu_\rho$ contains just the transformation of the field
$A_\mu$.

 Under an infinitesimal rotation
around the  $z$-axis with an angle $\delta\omega$, the
electromagnetic field $\vec A$ has a well defined transformation
rule, $A_i = A_i \rightarrow\epsilon_{ij3} A_j \delta\omega$,
which is independent of the origin of space coordinates. If one
considers the Noether term $\Xi^0_\rho$, Eq.(\ref{eq:intrinsec}),
an expression for the helicity is found:
\begin{equation}
S_z^{(i,i^\prime,p,\kappa,p^\prime,\kappa^\prime)} = (1/4\pi
c)\int_{\cal V}~d^3x ~ ({\vec E^{(i)}_{p,\kappa}} \times \vec
A^{(i^\prime)}_{p^\prime,\kappa^\prime})_z.
\end{equation}
 This is another dynamical constant for Weber EM modes as can
 be directly verified by substituting
the general expression for the vectors $\vec E$ and $\vec A$ in
terms of the Hertz parabolic modes. As expected, it turns out that
for a given  mode $p,\kappa$, the helicity $S_z$ is different from
zero only if the amplitudes ${\mathcal A}^{(TM)}$ and ${\mathcal
A}^{(TE)}$ that define the polarization of a mode are complex. The
Noether term $\Lambda_\rho^0$, Eq.(\ref{eq:extrinsec}), defines
the density of the $z$-component of the orbital angular momentum
\begin{equation}
 {\cal
L}_z^{(i,p,\kappa,p^\prime,\kappa^{\prime})}=\frac{1}{8\pi
c}\sum_j(\vec E^{(i)}_{p,\kappa})_j (u\partial_v-v\partial_u)
(\vec A^{(i)}_{p^\prime,\kappa^\prime})_j
\end{equation}
For Weber modes, similarly to periodic plane waves \cite{lenstra},
${\cal L}_z = \vec \nabla\cdot \vec G + \delta {\cal L}_z$ with
the latter term becoming zero for $p=p^\prime, \kappa
=\kappa^\prime$. Explicitly,
\begin{eqnarray}
\vec G &=&\frac{kh^2}{ 8\pi k^\prime c} \sum_j(h\vec
A^{(i)}_{p,\kappa})_j\vec {\mathbb M}(h\vec
A^{(i)}_{p^\prime,\kappa^\prime})_j \\
 \delta {\cal
L}^{(TE)}_z &=&\frac{k^2i}{8\pi k_z c} \sum_{j=u,v}(\partial_j h
\vec{\mathbb {N}} \Psi_{p,\kappa})\cdot(\partial_j h\vec{\mathbb
{M}}
\Psi_{p^\prime,\kappa^\prime}) \nonumber\\
 \delta {\cal
L}^{(TM)}_z &=&\frac{k_zk_z^\prime}{kk^\prime}\delta {\cal
L}^{(TE)}_z -\frac{k_\bot^2k_\bot^{\prime 2}h^4i}{8\pi k_z
c}(\vec{\mathbb M}\Psi_{p,\kappa})\cdot(\vec{\mathbb N}
\Psi_{p^\prime,\kappa^\prime})\nonumber
\end{eqnarray}
 Thus, a not null orbital angular
momentum for a given Weber mode $(p,\kappa)$ in a volume ${\cal
V}$ can be due just to a flux of the vector field $\vec G$ through
the boundary surface.

Let us consider ${\mathbb A}$ as generator of a transformation for
the vector field $\vec A$. Noether theorem as described above
concerns first order differential operators as generators of
continuous symmetries while $\mathbb A$ contains second order
operators. Nevertheless, the $\Xi^0_\rho$ term associated to this
transformation gives rise to
\begin{eqnarray}
\mathfrak{A}^{(i,p,\kappa,p^\prime,\kappa^{\prime})}=
\frac{1}{4\pi c} \int_{\cal V}  \sum_j (\vec E_{p,\kappa}^{(i)})_j
\mathbb{A}
(A_{p^\prime,\kappa^{\prime}}^{(i)})_j d^3x&&\nonumber\\
\sum_j (\vec E_{p,\kappa}^{(i)})_j \mathbb{A} (\vec
A_{p^\prime,\kappa^\prime}^{(i)})_j=
 k_{\perp} a  \vec{E}_{p,\kappa}\cdot \vec A_{p^\prime,\kappa^\prime}
 + \nabla \cdot \vec{C}^{(i)}&&\label{eq:barb}\\
 \vec C^{(TE)} = \vec C, \vec
C^{(TM)}=\frac{k_zk_z^\prime}{kk^\prime}\vec
 C, \vec C=-(\partial_{ct} \Psi_{p,\kappa})\vec{\mathbb{M}}
 \Psi_{p^\prime,\kappa^\prime}&.&\nonumber
 \end{eqnarray}
The first resulting term in Eq.~(\ref{eq:barb}) is proportional to
the integrand that defines the energy, Eq.~(\ref{eq:energy}).
Thus, for integrations  over a finite volume $\mathcal{V}$,  $\vec
C^{(i)}$ defines the flux of $\mathfrak{A}$ through the surface
around the integration volume.
 Eq.~(\ref{eq:barb}) supports the identification of
$ \mathfrak{A}$ as the electromagnetic dynamical variable related to the
generator $\mathbb{A}$.

As for the discrete symmetry, using the properties of the
scalar function $U$ under the reflection of $u$ and the expression
of the EM modes in terms of Hertz potentials, it is straightforward
to find the reflection properties of the electric field $\vec E_{p,\kappa}$ for each mode:
\begin{eqnarray}
{\mathfrak{P}}_u~\vec E^{(TE)}_{p,\kappa}&=&(-1)^p(-E^{(TE)}_{p,\kappa,x},
E^{(TE)}_{p,\kappa,y},
E^{(TE)}_{p,\kappa,z}),\nonumber\\
{\mathfrak{P}}_u~\vec E^{(TM)}_{p,\kappa}&=&(-1)^p(E^{(TM)}_{p,\kappa,x},
-E^{(TM)}_{p,\kappa,y},E^{(TM)}_{p,\kappa,z}).
\end{eqnarray}
\section{Quantization of the EM field in terms of Weber modes}
Standard quantization rules require a proper normalization of the
EM modes so that each travelling photon carries an energy
$\hbar\omega$. The classical electric field amplitude is
substituted by the electric field per photon times the creation
operator for the given travelling mode,
$\mathcal{A}_{\kappa}^{(i)} =  \varepsilon_{\kappa}
\hat{a}_{\kappa}^{(i)}$, $|\varepsilon_{\kappa}|^{2} = \hbar/ k^2
k_{\perp}^2$. The quantum energy and the momentum along $z$
operators take the form:
\begin{equation}
\hat {\cal E}=  \sum_{i,\kappa}\hbar\omega ~\hat{N}_{\kappa}^{(i)} ,
\qquad \hat {\cal P}_z=  \sum_{i,\kappa}\hbar k_z
~\hat{N}_{\kappa}^{(i)},
\end{equation}
in terms of the number operator:
\begin{equation}
\hat {N}_{\kappa}^{(i)} = \frac{1}{2} \Big(\hat a^{(i)\dagger}_{\kappa}
\hat a^{(i)}_{\kappa} + a^{(i)}_{\kappa} \hat a^{(i)\dagger}_{\kappa}
\Big),\quad
[a^{(i)}_{\kappa},a^{(j)\dagger}_{\kappa^\prime}] =
\delta_{i,j}\delta_{\kappa,\kappa^\prime},
\end{equation}
allowing the identification of $\hbar k_{z}$ and $\hbar \omega$ with
the photon momentum along $z$ and the photon energy respectively.
As for the helicity,
\begin{equation}
\hat{ S}_z =  \sum_{\kappa} \frac{i \hbar k_z c}{2\omega}
\big( \hat a^{(TE)\dagger}_{\kappa}   \hat a^{(TM)}_{\kappa} -  \hat a^{(TE)}_{\kappa} \hat a^{(TM)\dagger}_{\kappa} \big).
\end{equation}
A quantum analysis of the relation between polarization and helicity can be carried out in analogy with the study in
Ref.~\cite{Jauregui2005} for Bessel fields.
Finally, in the quantum realm the field operator associated to ${\mathbb A}$ is
\begin{equation}
\hat{\mathfrak{A}}
= \sum_{i,\kappa}  \hbar^2 k_{\perp} a ~\hat{N}_{\kappa}^{(i)} .
\end{equation}
An overall factor $\hbar$ was introduced so that the dynamical
variable $\hat{\mathfrak{A}}$ for a
photon has units of linear momentum times angular momentum as expected for
the
quantum variable associated to $p_{y}l_{z}$.

\section{Discussion}
The parabolic-cylindrical modes differ from other separable cylindrical modes by having
 $(l_zp_y + p_y l_z)/2$ as a symmetry operator. We showed that the quantum numbers of the
 EM modes $\{k_z,\omega,a \}$ are related to their linear momentum along $z$, the energy, and
 the symmetrized product of the angular momentum along $z$ and the momentum along $y$. The helicity, a property intrinsic
 to the vector nature of the EM field, was shown to be
 diagonal in the circular basis resulting from the complex superposition of TE and TM modes.
 The dynamical variable $\mathfrak{A}$ is gauge
 dependent although it can be written in a gauge independent
 looking form for monochromatic modes for which
 $\vec E_{p,\kappa} = i\omega \vec A_{p,\kappa}$. Contrary to
 standard EM dynamical variables which depend on products of $\vec E$ and $\vec B$,
 ${\mathfrak{A}}$ depends on the products of $\vec E$, $\vec B$, and their derivatives.
 Since Weber EM modes form a complete set,
 ${\mathfrak{A}}=(1/4\pi c)\sum_j \int_{\cal V}d^3xE_j {\mathbb A} A_j$
 will be a conserved quantity for any EM wave $\vec A$ whenever the flux of ${\mathfrak{A}}$
 through asymptotic parabolic cylinder surfaces at infinity is null. This flux can be evaluated
writing the given EM field $\vec A$ as a superposition of Weber
modes and applying Eq.~(\ref{eq:barb}).

 The mechanical effects of Weber beams on cold atoms deserve a detailed study
 both quantum mechanically and semiclassically (in complete analogy to that
 already done for Mathieu beams \cite{Rodriguez2008}). However, there are
 some qualitative features that can be expected without performing
 such an analysis. For instance, since under stationary conditions,
 cold noninteracting atoms in a red-detuned light beam  have a higher
 probability of being located in the higher intensity regions of the
 beam, the corresponding squared atomic wave function  mimics the
 intensity pattern of the light field. Thus, for a Weber lattice,
 the atomic wave function is expected to have a geometrical structure
 similar to that of the scalar Weber function, Eq.~(\ref{eq:total}).
 This structure gives rise to the eigenvalue equations Eqs.~(\ref{eq:par}-\ref{eq:A}).
 Thus, necessarily $\mathfrak{A}$ define a natural dynamical variable for
 the mechanical description of the $atomic$ cloud in a Weber lattice.
 A careful analysis concerning this idea is in progress.


\end{document}